\documentclass[12pt,manuscript]{aastex}
\usepackage{amsmath}
\usepackage{epstopdf}
\usepackage{multirow}
\usepackage{lineno}
\usepackage{subfigure}
\usepackage{rotating}
\usepackage{longtable}
\linenumbers*[1]

\newcommand{\msun}{M$_{\odot}$}

\slugcomment{Accepted to {\it The Astrophysical Journal}}

\shorttitle{BIASES IN PLANET TRANSIT SEARCHES}
\shortauthors{Gaidos \& Mann}

\begin{document}


\title{Objects in {\it Kepler's} Mirror May be Larger Than They
  Appear:\\ Bias and Selection Effects in Transiting Planet Surveys}

\author{Eric Gaidos} 
\affil{Department of Geology and Geophysics, University of Hawai`i at M\={a}noa, Honolulu, HI 96822}

\email{gaidos@hawaii.edu}

\and

\author{Andrew W. Mann}
\affil{Institute for Astronomy, University of Hawai`i at M\={a}noa, Honolulu, HI 96822}

\begin{abstract}
  Statistical analyses of large surveys for transiting planets such as
  the {\it Kepler} mission must account for systematic errors and
  biases.  Transit detection depends not only on the planet's radius
  and orbital period, but also on host star properties.  Thus, a
  sample of stars with transiting planets may not accurately represent
  the target population.  Moreover, targets are selected using
  criteria such as a limiting apparent magnitude.  These selection
  effects, combined with uncertainties in stellar radius, lead to
  biases in the properties of transiting planets and their host stars.
  We quantify possible biases in the {\it Kepler} survey.  First,
  Eddington bias produced by a steep planet radius distribution and
  uncertainties in stellar radius results in a 15-20\% overestimate of
  planet occurrence.  Second, the magnitude limit of the {\it Kepler}
  target catalog induces Malmquist bias towards large, more luminous
  stars and underestimation of the radii of about one third of
  candidate planets, especially those larger than Neptune.  Third,
  because metal-poor stars are smaller, stars with detected planets
  will be very slightly ($<$0.02 dex) more metal-poor than the target
  average.  Fourth, uncertainties in stellar radii produce correlated
  errors in planet radius and stellar irradiation.  A previous
  finding, that highly-irradiated giant are more likely to have
  ``inflated'' radii, remains significant, even accounting for this
  effect.  In contrast, transit depth is negatively correlated with
  stellar metallicity even in the absence of any intrinsic
  correlation, and a previous claim of a negative correlation between
  giant planet transit depth and stellar metallicity is probably an
  artifact.
\end{abstract}


\section{Introduction \label{sec.introduction}}

When an exoplanet's orbital plane lies along our line of sight, the
planet will transit its host star, periodically obscuring a small
portion of the stellar disk and producing detectable dips in a
photometric lightcurve.  The first transits (of a planet previously
discovered by Doppler) were observed in 1999
\citep{Henry2000,Charbonneau2000}.  The first exoplanet to be detected
with the transit technique was confirmed by Doppler observations in
2002 \citep{Konacki2003}.  As of 18 October 2012, 288 confirmed
transiting planets in 233 systems have been reported
\citep{Schneider2011}.  The {\it Kepler} mission, operating since
2009, has identified more than 2300 candidate transiting planets ({\it
  Kepler} Objects of Interest or KOIs) \citep[hereafter
B12]{Batalha2012}.  Although only a small fraction of KOIs have been
confirmed, the false positive rate is thought to be low
\citep{Morton2011a,Lissauer2012}, but see \citet{Santerne2012}.

Transit searches are more sensitive than Doppler searches to the
smallest planets because the transit signal scales with the square of
the planet radius $R_p$, while the Doppler signal of a rocky planet
scales approximately as $R_p^4$ \citep{Valencia2007b}.  {\it Kepler}
has already discovered 90 candidates possibly {\it smaller} than
Earth.  Transiting planets are of special interest because their radii
can be estimated from the transit signal.  If the transit is not
grazing, the fractional decrease $\delta$ in the star's observed flux
is
\begin{equation}
\delta = \left(\frac{R_p}{R_*}\right)^2,
\label{eqn.depth}
\end{equation}
where $R_*$ is the stellar radius.  A measurement of $\delta$ combined
with knowledge of $R_*$ yields the planet radius.  Because the
inclination of a transiting planet's orbit is near 90$^{\circ}$, the
mass of the planet can also be unambiguously established from Doppler
observations.  Combinations of mass and radius can be compared with
predictions by models of interior structure
\citep{Seager2007,Grasset2009a,Rogers2011}.  Spectroscopy or
spectrophotometry during transits can detect or rule out constituents
in an atmosphere \citep{Charbonneau2002,Bean2010,Desert2011}, and
secondary eclipses (occultations of the planet) can constrain
temperature and albedo \citep{Charbonneau2005,Knutson2008,Rowe2008}.
Additional planets can be discovered by variation in transit times
\citep{Agol2005,Forda}.

Analyses of large samples of transiting planets, including the catalog
of KOIs, have attempted to ascertain properties of transiting planet
populations, e.g., whether they are segregated into discrete groups
\citep{Fressin2009a}, the distribution with planet radius
\citep{Howard2012} the dependence of planet occurrence on the
metallicity of the host star \citep{Schlaufman2011,Buchhave2012}, the
effect of stellar irradiance on giant planet radius
\citep{Demory2011a,Enoch2012}, and the occurrence of planets compared
to Doppler surveys \citep{Gaidos2012,Wolfgang2012,Wright2012}.  In the
case of {\it Kepler}, lack of Doppler confirmation for most candidate
planets as well as detailed spectroscopic characterization of the
stars make it important to properly account for any systematic effects
or biases.

Detection of a planet in a transit survey depends on the properties of
the planet, most notably $R_p$ (Equation \ref{eqn.depth}), but also on the
orbital period because it determines the number of transits that are
observed and the total transit signal.  \citet{Gaudi2005},
\citet{Gaudi2005a}, and \citet{Pont2006a} pointed out that
transit-selected samples are biased toward large planets on
short-period orbits.  These biases can be extreme in ground-based
surveys which suffer correlated (``red'') noise from variations in
atmospheric transmission and discontinuous observing windows.

Equation \ref{eqn.depth} also shows that planets of a given radius will be
more readily detected around smaller stars.  This has, in part,
motivated transit searches for planets around M dwarf stars
\citep{Tarter2007,Gaidos2007,Charbonneau2007b}.  In this case, a
property of the host star, as opposed to the planet, influences the
likelihood that a transiting planet will be detected, and that both
star and planet will be included in a transit-selected sample.  Thus a
selection effect will act on stellar radius, or on any property that
is related to stellar radius, such as metallicity.  This will produce
systematic offsets or biases in the properties of stars hosting known
transiting planets relative to the properties of the target sample.

The construction of a target catalog itself can also produce selection
effects in a transit survey.  Most notable among constraints on target
stars is an apparent magnitude limit because of a signal-to-noise
ratio (SNR) requirement, or the need to confirm candidate transiting
systems using Doppler observations.  A magnitude limit will cause
(Malmquist) bias towards more luminous stars; these can be included to
larger distances and hence sample a larger volume of space.  At a
given effective temperature, luminosity is uniquely related to stellar
radius, and hence this is also a bias towards larger stars that,
unmitigated, will affect the detection of planets and estimates of
their radii.

Some of these effects would disappear or could be corrected if stellar
parameters, i.e. radius, were precisely established.  But, up to now,
the large scale of transit surveys ($10^4-10^7$ stars) has precluded
such determinations.  Neither radius nor mass are directly observable
for distant, single stars such as {\it Kepler} targets.  The
properties of most {\it Kepler} stars have been inferred by comparing
stellar models to the broad-band photometry of the {\it Kepler} Input
Catalog (KIC) \citep[hereafter Br11]{Brown2011a}.  Few spectra and
almost no parallax (distance) measurements are available, and most
stars have only upper limits on proper motion.  KIC estimates of
stellar radii have large uncertainties due to (i) errors in
photometry; (ii) degeneracies between stellar parameters and colors;
and (iii) errors in the models themselves.  While KIC estimates of
stellar effective temperature are comparatively robust, those of
surface gravity ($\log g$) and metallicity (Fe/H) are not as reliable
(Br11).  Br11 concluded that no gravity or radius information could be
inferred for stars hotter than 5400~K ($g-r<0.65$).
\citet{Verner2011} found that the KIC and astroseismic radii of 500
solar-type stars have random discrepancies of order 50\% and a
systematic offset of about the same amount.  \citet{Bruntt2012} found
a similar scatter but negligible systematic offset in $\log g$ (and
hence the radius) of 93 solar-type {\it Kepler} stars.
\citet{Mann2012} found that many M-type stars that were classified as
dwarfs or were unclassified in the KIC are actually evolved stars.

Selection effects acting on uncertainties in stellar radius will bias
the observed properties of planet-hosting stars with respect to their
true distributions.  For example, while essentially all M-type hosts
of KOIs are {\it bona fida} dwarf stars \citep{Muirhead2012}, the vast
majority of the bright ($K_p < 14$) targets and some fainter stars are
giants \citep{Mann2012}.  This disparity is a result of the strong
selection effect on stellar radius described above; planets are far
more difficult to detect around giant stars due to their large size
and higher variability \citep{Huber2010a}.  Because of the relation
between planet radius and stellar radius (Equation \ref{eqn.depth}),
estimates of planet radius will likewise be affected.

Here, we quantify five effects produced by selection bias and
uncertainties in stellar parameters in the {\it Kepler} survey.  In
Section \ref{sec.scaling} we derive useful scaling relationships for
selection effects on transiting planet detection and target star
selection.  In Section \ref{sec.kepler} we apply these concepts to the
{\it Kepler} survey using the KOI catalog, parameters from the KIC,
and models of stellar evolution and stellar populations.  We describe
our methods and models in Section \ref{sec.methods}.  In Section
\ref{sec.eddington}, we calculate the effect of Eddington bias on the
radius distribution of KOIs as a result of uncertainties in stellar
radius.  In Section \ref{sec.malmquist} we describe the effect of
Malmquist bias on the magnitude-limited {\it Kepler} target catalog
and the preferential inclusion of more luminous, larger stars, thus
biasing downwards the radius of some KOIs.  In Section
\ref{sec.metallicity} we estimate the bias towards lower metallicity
among KOI-hosting stars as a consequence of the relationship between
stellar metallicity and radius on the main sequence.  In Section
\ref{sec.inflated} we describe how uncertainties in stellar radius
produce correlated errors in planet radius and stellar luminosity,
potentially affecting statistics describing the relationship between
``inflated'' giant planets and stellar irradiation.  In Section
\ref{sec.shrunken} we consider the effect of stellar metallicity on
transit depth and the interpretation of any correlation between
metallicity and the radii of giant planets.  We summarize our results
and describe current and future efforts to better determine the
parameters of {\it Kepler} stars in Section \ref{sec.discussion}.

\section{Analytical scaling relations \label{sec.scaling}}

In a transit survey, selection bias acts on a quantity $X$ (a stellar
or planetary parameter) when the probability $f$ that a star is
included in the survey, or that a planet is detected transiting a
star, depend on that parameter.  This bias is superposed on any real
correlations and will persist to the extent that the values of the
parameter and its effect on inclusion or detection are imperfectly
quantified.  The bias $\delta X$ is the difference between the
observed mean $\langle Xf \rangle/\langle f \rangle$ and the
intrinsic mean $\langle X \rangle$, or
\begin{equation}
\label{eqn.bias}
\delta X = \frac{\langle Xf \rangle - \langle X \rangle \langle f \rangle}{\langle f \rangle},
\end{equation}
where the brackets represent marginalizing over the population of
stars, subject to any constraints.  To derive useful scaling
relations, we chose apparent brightness (magnitude) and effective
temperature $T_e$ as independently varying parameters.  The first
fixes the noise level against which a transit must be detected.
Morever, the {\it Kepler} target catalog is magnitude-limited
\citep{Batalha2010}.  Among main sequence stars, $T_e$ is closely
related to mass, an important parameter of planet populations
\citep{Johnson2010,Howard2012}.  Unlike other stellar parameters, it
can be robustly estimated from KIC photometry
\citep[Br11,][]{Pinsonneault2012}).  Effective temperature is thus a
convenient plotting parameter which minimizes complications due to
variation in the planet population along the main sequence.
Nevertheless, values of $T_e$ do {\it not} map to unique values of
stellar mass because stars have different ages and metallicities and
plots with $T_e$ the dependence on mass should be considered
``blurred''.  Calculations using stellar models, as described below,
explicitly take into account the effects of age and metallicity.

\subsection{Selection bias due to transit detection \label{sec.transitbias}}

We first estimate the probability $f$ that a star is included in a
catalog of transiting systems.  The probability of detecting a planet
is calculated as a function of both stellar properties (radius and
mass $R_*$ and $M_*$) and planet properties (radius $R_p$ and orbital
period $P$), and then marginalized over planet properties using an
appropriate distribution function.  This yields $f$ as a function of
$R_*$ and $M_*$.  Equation \ref{eqn.bias} can then evaluated using a
stellar model that describes the intrinsic distributions of these
parameters and their relations to other observables.  In many
instances we can use scaling relations rather than exact relations
becaue Equation \ref{eqn.bias} is normalized.

We adopted a double power-law for the intrinsic distribution $f'$ of
planets with radius and orbital period
\citep{Cumming2008,Howard2010a,Mayor2011,Howard2012}:
\begin{equation}
\label{eqn.distribution}
df' \sim R_p^{-\alpha}P^{-\beta} d \ln R_p d \ln P,
\end{equation}
for $P$ larger than some minimum value $P_{min}$ where planets are
found.  Transit detection depends on the geometric probability that
the planet is on a transiting orbit, as well as the the signal (depth) of the transit relative to noise.  

In the absence of coherent or ``red'' noise from the atmosphere, the
signal-to-noise ratio of a single transit is
$\sqrt{N}\left(R_P/R_*\right)^2$, where $N$ is the total number of
photons detected during the event.  In an observation interval $t$
about $t/P$ transits will be observed, bringing the total number of
photons to $\sim Nt/P$.  Therefore the signal-to-noise ratio of the
co-added transits is
\begin{equation}
\label{eqn.snr0}
{\rm SNR} = \sqrt{\frac{Nt}{P}}\left(\frac{R_p}{R_*}\right)^2.
\end{equation}
At a given apparent brightness, $N \sim \tau$, where $\tau$ is the
transit duration.  When the transit impact parameter is low and the
transit chord is close to the stellar diameter, $\tau \approx 2R_*/V$.
Assuming a near-circular orbit, the transverse velocity $V$ can be
expressed in terms of the orbital period and mass of the star and
\begin{equation}
\tau \approx 2R_*\left(\frac{P}{2\pi GM_*} \right)^{1/3},
\end{equation}
where $G$ is the gravitational constant.  We derive a scaling relation
between SNR and planet/star properties by substituting $\tau$ for $N$
in Equation \ref{eqn.snr0} and ignoring constant factors:
\begin{equation}
\label{eqn.snr}
{\rm SNR} \sim R_p^2P^{-1/3}R_*^{-3/2}M_*^{-1/6}.
\end{equation}
Solving for $R_p$ gives a scaling relation for the radius of the
smallest planet on a given orbital period that can be detected at a
fixed SNR threshold:
\begin{equation}
\label{eqn.smallest}
R_{min} \sim R_*^{3/4}M_*^{1/12}P^{1/6}.
\end{equation}
Likewise, there is a relation for the maximum orbital period at which a planet of
a given radius $R_p$ can be detected at a fixed SNR threshold:
\begin{equation}
\label{eqn.pmax}
P_{max} \sim R_p^6 R_*^{-9/2} M_*^{-1/2}.
\end{equation}
$P_{max}$ is a sensitive function of $R_p$, underscoring why transit
surveys are highly biased towards the largest planets
\citep{Gaudi2005}.

To obtain the observed distribution $f$ of planets with $R_*$ and
$M_*$, we multiply the intrinsic distribution
(Equation \ref{eqn.distribution}) by the geometric probability that a
planets is on a transiting orbit.  For circular orbits this is
proportional to the ratio of the stellar radius to orbital semimajor
axis $R_*/a$ which, based on Newtonian orbital dynamics, scales as
$R_*M_*^{-1/3}P^{-2/3}$.  The observed planet distribution is:
\begin{equation}
\label{eqn.observed}
df \sim R_*M_*^{-1/3}R_p^{-\alpha}P^{-(\beta+2/3)} d \ln R_p d \ln P,
\end{equation}
We marginalize Equation \ref{eqn.observed} over both $P$ and $R_p$, first
integrating from $P_{min}$ to $P_{max}$.  The maximum period is also
limited by the observing window and the requirement that more than one
transit must be observed, e.g. $P_{max} \le t/3$.  

Integration of $P^{-(\beta+2/3)} d \ln p$ in Equation
\ref{eqn.observed} yields a factor proportional to $P_{min}^{-(\beta +
  2/3)} - P_{max}^{-(\beta + 2/3)}$.  If $P_{min} \ll t$, then
Equation \ref{eqn.pmax}, is used to re-express this as
$P_{min}^{-(\beta + 2/3)}\left[1-(R_p/R_{m})^{-(6\beta+4)}\right]$,
where $R_{m}$ is the radius of the smallest planet that can be
detected at $P = P_{min}$, i.e. that can be detected at all).
Integration of Equation \ref{eqn.observed} over $R_P$ from $R_{m}$ to
$\infty$ produces:
\begin{equation}
\label{eqn.observed2}
\int  df \sim R_*M_*^{-1/3}R_{m}^{-\alpha} P_{min}^{-(\beta+2/3)} \int_1^{\infty} x^{-(\alpha+1)}\left(1-x^{-(6\beta + 4)}\right) dx,
\end{equation}
were $x \equiv R_p/R_{m}$.  Because the $P_{min}$ factor and the
integral depend only on $\alpha$ and $\beta$, which are planet
population parameters and not stellar properties, they can be ignored
when calculating biases in stellar properties.  Substituting for
$R_{m}$ and retaining only factors that depend on stellar properties,
\begin{equation}
\label{eqn.occur}
f \sim R_*^{1-3\alpha/4}M_*^{-(1/3+\alpha/12)}
\end{equation}
All else being equal, planets are more likely to be detected around
stars with smaller radii (because transit depths are larger) and/or
lower masses (because transit durations are longer). Smaller stars are
thus more likely to appear in a transit-selected sample.  In the case
of mass-radius relation $R_* \sim M_*^{0.8}$ for zero-age solar-type
stars \citep{Cox2000} and a planet radius distibution power-law index
$\alpha = 2.6$ \citep{Howard2012}, then $f \sim M_*^{-1.31}$.  This is
simply a statement that smaller (and more) planets can be detected
around lower mass stars.  Older stars will have a steeper mass-radius
relation, and as a result the dependence of $f$ on $M_*$ will be more
pronounced.

At a given apparent brightness (observed flux), the quantity
$BR_*^2/d^2$ is fixed, where $B$ is the stellar surface brightness in
the bandpass of interest and $d$ is the distance to the star.
Substituting, $R_* \sim d/\sqrt{B}$ into Equation \ref{eqn.occur}, and
assuming that the planet population is distance-independent so that
the distance factor can be moved outside the period and radius
integrals, the scaling relation for observed occurrence becomes:
\begin{equation}
\label{eqn.occur2}
f \sim d^{1-3\alpha/4}B^{3\alpha/8-1/2}M_*^{-(1/3+\alpha/12)}.
\end{equation}
If $\alpha > 4/3$, \citep[][e.g.]{Howard2012}, closer and hotter host
stars are more likely to be included in transit-selected samples.
Stellar age and metallicity, which affect the relationships between
stellar mass, radius, and surface brightness, are also biased as a
result.  A correlation between stellar properties and distance can
modulate the degree of this bias.  For example, if more distant stars
tend to be more evolved along the main sequence and thus hotter, the
bias will be less than if $d$ and $B$ are independent.  Equation
\ref{eqn.occur2} does not consider that star of a certain mass or
radius may be {\it over-represented} in the parent population: this is
discussed in the next section.

\subsection{Selection bias due to target selection \label{sec.targetbias}}

The target catalogs of transit surveys such as {\it Kepler} are
selected using a number of criteria, and chief among these is apparent
magnitude.  A magnitude-limited sample of stars will be biased towards
the most luminous objects, which will be included to greater distances
\citep{Malmquist1922}.  These stars may be either more massive, more
evolved, or both.  At a given $T_e$ and thus fixed surface brightness
$B$ (ignoring the weak dependence of surface brightness on gravity and
metallicity), the signal $N$ from a star during a transit will scale
as $(R_*/d)^2$.  Modifying Equation \ref{eqn.snr0} appropriately, we
find that the transit signal-to-noise ratio scales as
\begin{equation}
\label{eqn.snr2}
{\rm SNR} \sim R_p^2P^{-1/3}R_*^{-1/2}M_*^{-1/6}d^{-1}.
\end{equation}
The smallest planet that can be detected at a given SNR will scale as
\begin{equation}
\label{eqn.small}
R_{small} \sim P^{1/6}R_*^{1/4}M_*^{1/12}d^{1/2}.  
\end{equation}
Multiplying a power-law distribution of planet radii
(Equation \ref{eqn.distribution}) by the probability that a planet is on a
transiting orbit ($\sim M_*^{-1/3}R_*$) and integrating over all
planet radii down to $R_{small}$ gives the relation
\begin{equation}
\label{eqn.number}
f \sim P^{-\alpha/6}R_*^{1-\alpha/4}M_*^{-(1/3+\alpha/12)}d^{-\alpha/2}.
\end{equation}
At a fixed color/temperature/surface brightness $B$, a
magnitude-limited survey will include stars of radius $R_*$ out to a
distance $d_{max} \sim R_*$.  Assuming, for the moment, that transits
can be detected to arbitrarily large distances, then integrating
Equation \ref{eqn.number} over a homogeneous volume of radius
$d_{max}$ yields
\begin{equation}
  f \sim P^{-\alpha/6}R_*^{4-3\alpha/4}M_*^{-(1/3+\alpha/12)}.
\end{equation}
For $\alpha = 2.6$ and at a given $P$, $f$ scales as
$R_*^{2.05}M_*^{-0.55}$.  This relation illustrates how larger, more
evolved stars can be preferentially included in a transit-selected
sample despite the fact that transits of these stars are more
difficult to detect.

Although target stars in a magnitude-limited sample will be included
to a distance $d_{max} \sim R_*$, a planet of radius $R_p$ can only be
detected to a distance $d_{det}$ where, according to
Equation \ref{eqn.snr2},
\begin{equation}
d_{det} \sim R_p^2P^{-1/3}R_*^{-1/2}M_*^{-1/6}.
\end{equation}
The detection limit decreases with $R_*$ while the inclusion limit
$d_{max}$ {\it increases} with $R_*$.  These limits coincide ($d_{max}
= d_{det}$) at a stellar radius $\tilde{R_*}$: 
\begin{equation}
\label{eqn.complete}
\tilde{R_*} = \tilde{R_0}R_P^{4/3}P^{-2/9}M_*^{-1/9},
\end{equation}
where $\tilde{R_0}$ is a constant factor, $P$ is in days, $R_P$ is in
Earth radii and $M_*$ is in solar masses.  (We calculate values of of
$\tilde{R_0}$ for the {\it Kepler} survey in Section
\ref{sec.malmquist}.)  Detections of planets of a given size around
stars with $R_* < \tilde{R_*}$ is magnitude-limited and subject to a
stellar radius bias that scales as $\sim R_*^{4}$, because the sample
volume increases as $R_*^3$ and the transit probability increases as
$R_*$.  For stars with $R_* > \tilde{R_*}$, a survey is limited to a
volume propoortional to $d_{det}^{3} \sim R_*^{-3/2}$ (see Equation
\ref{eqn.small}), and the bias scales as $\sim R_*^{-1/2}$, a weak
dependence on $R_*$ in the opposite sense.  The critical stellar
radius $\tilde{R_*}$ is most sensitive to planet radius and the
dependence on period and stellar mass is weak.

\section{Application to the {\it Kepler} transit survey \label{sec.kepler}}

\subsection{Methods \label{sec.methods}}

To evaluate biases and selection effects in the {\it Kepler} survey we
modeled target stars with isochrones from the Dartmouth Stellar
Evolution Database \citep{Dotter2008} interpolated onto a 0.1-dex grid
of metallicities using the on-line tool.  For each star, we compared
adjusted KIC parameters ($T \equiv T_e$, ${\mathcal G} \equiv \log g$,
${\mathcal F} \equiv$ [Fe/H]) to model predictions using Bayesian
statistics.  Specifically, we calculated a probability or weight $w$
for each model according to:
\begin{equation}
\label{eqn.weight}
w = e^{-\left[\frac{(T - \hat{T})^2}{2\sigma_T^2} +\frac{({\mathcal G} - \hat{\mathcal G})^2}{2\sigma_{\mathcal G}^2} + \frac{({\mathcal F} - \hat{\mathcal F})^2}{2\sigma_{{\mathcal F}}^2}\right]}p(M_*)p(t_*)p({\mathcal F})p(\zeta),
\end{equation}
where parameters with a ``hat'' are the Dartmouth model values and
$p(M_*)$, $p(t_*)$, $p({\mathcal F})$, and $p(\zeta)$ are the priors
for initial stellar mass (initial mass function, IMF), age,
metallicity, and a modified distance modulus $\zeta \equiv \mu + 5
\log_{10} \sin b$, where $b$ is the galactic latitude.  The modified
distance modulus accounts for the finite dispersion of stars above the
plane of the Milky Way, but neglects the vertical displacement of the
Sun.  We used an SDSS $r$-band modulus $\mu = m_r - M_r$, where $m_r$
is the observed apparent magnitude and $M_r$ is the absolute magnitude
from the Dartmouth models.  We ignored interstellar extinction, which
will be $<0.5$ magnitudes \citep{Schlegel1998}.  (While estimation of
stellar parameters is sensitive to interstellar {\it reddening}, the
amount of interstellar {\it extinction} is small compared to the
uncertainties in the distance modulus.)  Estimates of $T_e$ and [Fe/H]
from the KIC were adjusted by -100~K and 0.17 dex, respectively and we
used $\sigma_{T_e} = 200$~K, $\sigma_{\log g} = 0.36$~dex, and
$\sigma_{{\rm Fe}} = 0.3$~dex, based on a comparison of KIC values
with those spectroscopic values listed in B12.

For priors we adopted the \citet{Kroupa2002} IMF, and a uniform age
distribution over 1-13 Gyr.  The latter corresponds to a constant rate
of star formation since the advent of the galactic disk
\citep{Oswalt1996,Liu2000}, but ignores the youngest stars, around
which planets are more difficult to detect.  The metallicity
distribution of {\it Kepler} target stars is unknown and may be
complex; the field is not parallel to the Galactic plane and may
include members of a metal-poor ``thick disk'' population
\citep{Ruchti2011}.  We used the metallicity distribution predicted by
the the TRILEGAL stellar population model \citep{Vanhollebeke2009} as
a prior.  Stars in the direction of the center of the {\it Kepler}
field ($\ell = 76.32^{\circ}$, $b = 13.5^{\circ}$) were simulated to a
cutoff magnitude $K_P = 16$.  When compared to 2MASS counts, TRILEGAL
counts agree with observations at least down to $b=10^{\circ}$, but
fail at $b=0$, possibly due to incorrectly modeled bulge red giant
branch stars and dust \citep{Girardi2005}.  However, the Kepler field
cuts off at $b=6^{\circ}$ and and only 18 of the 84 CCD centers lie at
$b<10^{\circ}$.  The (mostly default) values for TRILEGAL parameters
are listed in Table \ref{tab.params}.

TRILEGAL also reports a value of $\mu$ for each simulated star and we
used these to construct a prior distribution of $\zeta$.  Our priors
are relaxed in the sense they only exclude very unlikely masses, ages,
or metallicities.  It is also possible to impose priors on the stellar
parameters $T_e$ and $\log g$ using the predictions of a stellar
population model, but we consider such predictions too uncertain to
justify this approach.

\begin{figure}
\epsscale{0.7}
\plotone{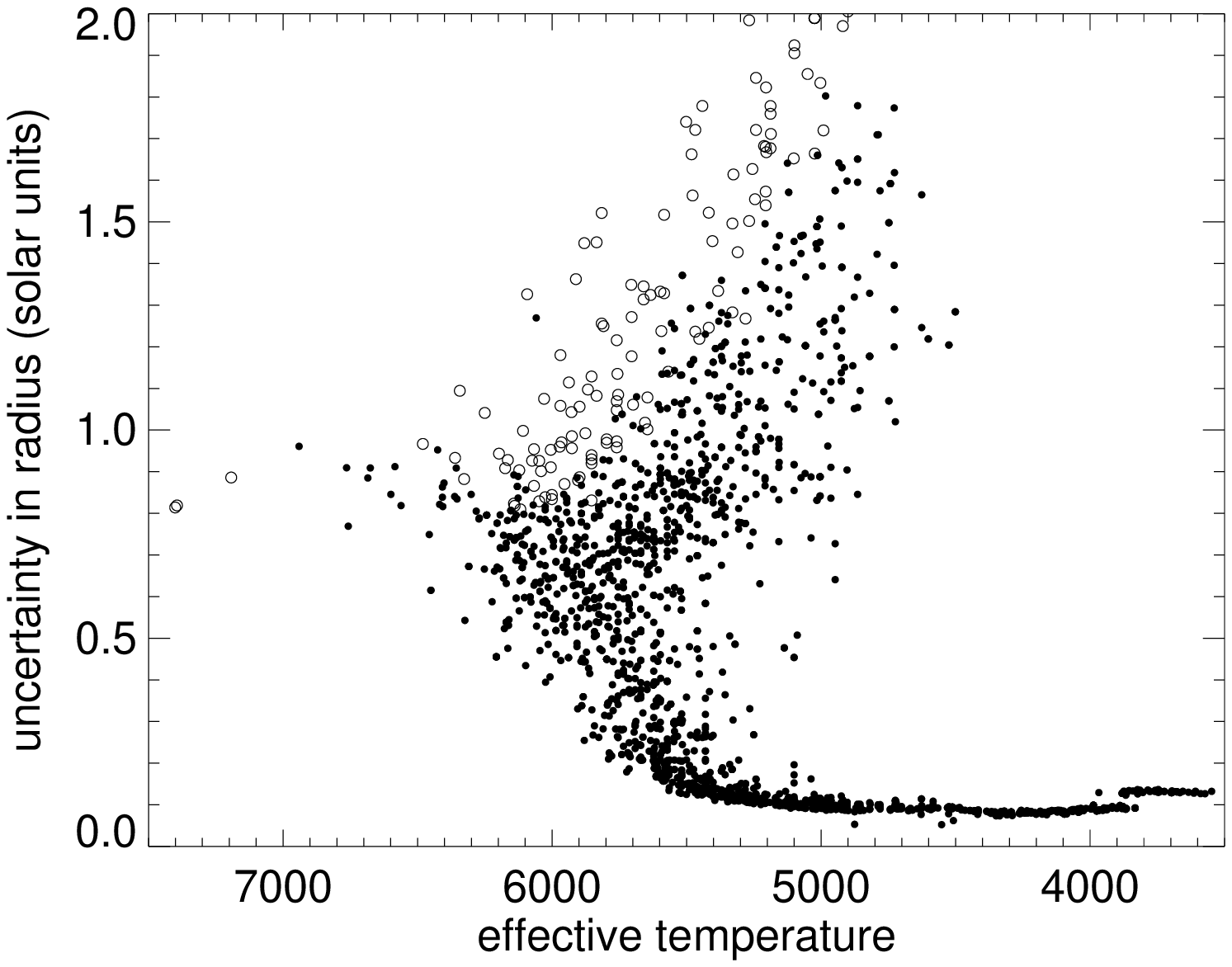}
\caption{Uncertainty in radius of stars hosting KOIs, defined as one
  half of the range encompassing 68\% of the probability distribution
  of radii.  A few stars with the largest uncertainy are off-scale.
  As defined, the uncertainty can exceed the mean or most likely value
  and does in some cases.  The adjusted KIC effective temperature is
  plotted on the abscissa.  Solid points have KIC $\log g >4$
  (``dwarfs''), while open points have $\log g <4$ (``giants'').
  While K-type dwarfs have uncertainties of as little as $\sim$15\%,
  the radius of F- and many G-type stars is uncertain by $\ge 100$\%
  because of the proximity of the giant and dwarf branches.  The
  discontinuity at $T_e \sim 3900$~K is an artifact of the grid of
  models and the sensitivity to very large M
  giants.  \label{fig.radunc}}
\end{figure}

For each star, Equation \ref{eqn.weight} returns an array of values
for $w$ corresponding to the grid of Dartmouth models.  Most values of
$w$ are negligibly small and the corresponding models were ignored.
From the remainder, the most probable (highest $w$) model and
accompanying parameters such as $R_*$ were identified.  Statistics of
the distribution of possible values were calculated, e.g:
\begin{equation}
  \bar{R_*} = \frac{\sum_i w_iR_*(i)}{\sum_i w_i}.
\end{equation}
Because the distributions can be very non-gaussian, we defined the
fractional uncertainty in a stellar parameter to be one-half of the
range encompassing 68\% of the total probability (normalized $w$)
divided by the most probable value.  We found that uncertainties in
the radii of late G- and K-type dwarf stars hosting KOIs is typically
$\sim$15\%, but are substantialy higher ($\sim100$\%) among some F-
and G-type stars because of the coincidence of the dwarf and
(sub)giant branches (Figure \ref{fig.radunc}).  Evolved stars
(i.e. KIC $\log g < 4$) also have comparatively larger uncertainties.
The cluster of putative M ``dwarfs'' with radius uncertainties of
$\sim$25\% might be misclassified giant stars \citep{Mann2012}.  Our
estimated uncertainties are certainly lower bounds because (1) the
errors in the stellar parameters $T_e$, [Fe/H], and especially $\log
g$ are themselves not gaussian-distributed, as presumed in Equation
\ref{eqn.weight}; and (2) we do not consider errors in the Dartmouth
models themselves.

\subsection{Eddington Bias \label{sec.eddington}}

Eddington bias occurs when errors in measurement scatter more frequent
values in a population to less frequent values at a higher rate than
the reverse process.  This systematically inflates the observed
frequency of rare members \citep{Eddington1913}.  Because the
distribution of planets with radius is a steep power law
\citep{Cumming2008,Howard2010a,Mayor2011,Howard2012}, errors in radius
(fractional standard deviation $\sigma_R$) will bias the number of
larger planets upwards.  This will inflate the rate of planet
occurrence $f$ above a given cutoff in radius $R_C$.  Planets with
radius $R_P$ will appear to be larger than the cutoff if the error in
stellar radius is larger than $R_C/R_P - 1$.  If errors in stellar
radius are gaussian-distributed, the fraction of stars that satisfy
that condition is ${\rm
  erfc}\left((R_C/R_P-1)/(\sqrt{2}\sigma_R)\right)/2$.  The fractional
upward bias in planet occurrence is the integral of this function with
the normalized planet radius distribution, minus the intrinsic
occurrence (normalized to unity):
\begin{equation}
\label{eqn.eddington}
\Delta f = \frac{\alpha}{2} \int_0^{\infty}x^{-(\alpha+1)}{\rm erfc} \left(\frac{x^{-1}-1}{\sqrt{2}\sigma_R}\right)dx  - 1,
\end{equation}
where $x \equiv R_p/R_C$.  $\Delta f$ increases with $\sigma_R$ and, if
$\alpha = 2.6$, reaches 18\% when $\sigma_R=30$\%.

\begin{figure}
\epsscale{0.7}
\plotone{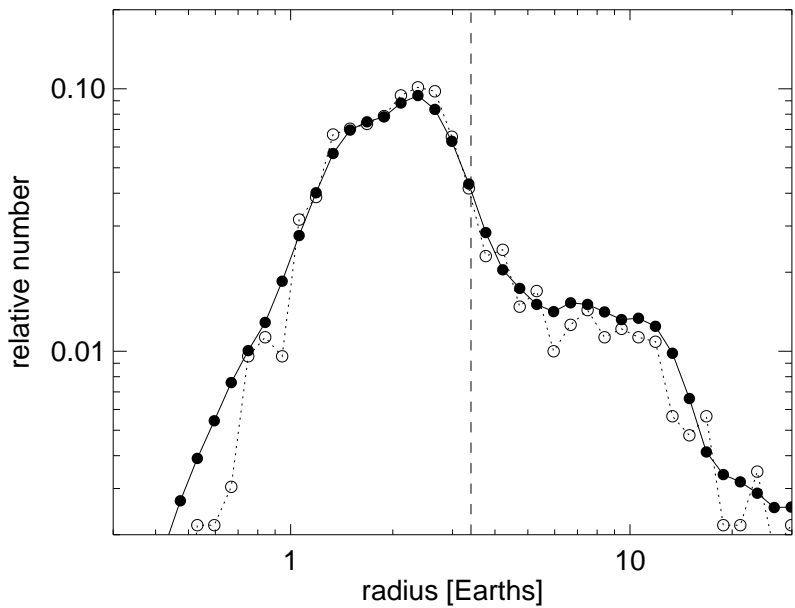}
\caption{Observed (uncorrected) distribution of KOI radii (open points
  and dashed line), and a distribution simulating the effects of
  Eddington bias (filled points and solid line).  The latter is
  constructed by adjusting the ratio of each planet candidate by the
  ratio of the most likely stellar radius to every possible radius
  among stellar models, weighted by a likelihood factor
  (Equation \ref{eqn.weight}).  The two normalized distributions are equal
  at $R_p = 3.4R_{\oplus}$.  The biased distribution has a shallower
  slope at small radii, a less pronounced bump at Jupiter-size, and a
  higher occurrence of planets larger than the completeness
  limit.  \label{fig.eddington}}
\end{figure}

We estimated the amount of Eddington bias in the apparent radius
distribution of KOIs using the procedures described in Section
\ref{sec.methods}.  For each KOI we calculated the likelihood weight
$w$ (Equation \ref{eqn.weight}) for all possible stellar models
consistent using the parameters of the host star.  Corresponding to
each model we calculated a revised planet radius $R_p \times
(R'_*/R_*)$, where $R_p$ is the radius from B12, $R'_*$ is the model
stellar radius and $R_*$ is the stellar radius of the maximum
likelihood model (highest $w$).  The radius distribution, weighted by
$w$, is summed over all KOI host stars and normalized.  This is
compared to the observed distribution of $R_p$ (Figure
\ref{fig.eddington}).  The latter is not the {\it intrinsic}
distribution, which must account for the probability that a planet
transits and is detected \citep{Howard2012}.  As expected, Eddington
bias increases the apparent number of Neptune-size and larger planets.
The bias is 17\% above $3.4R_{\oplus}$, demarcated by the vertical
dashed line in Figure \ref{fig.eddington}, where the normalized
distributions are equal.  The bias also suppresses the peak in the
distribution at a Jupiter radius.  Corollaries of these results are
that the actual occurrence rate of Neptune-size planets is smaller
than previously reported \citep[][i.e.]{Howard2012}, and that the
intrinsic peak at $R_p \sim 1R_J$ is more pronounced than is apparent.

In addition, Eddington bias decreases the apparent slope in the radius
distribution (Figure \ref{fig.eddington}).  This is a consequence of
the observed turnover in the number of planets smaller than
$\sim2R_{\oplus}$, and whether more large planets are scattered to
smaller radii than vice versa.  {\it Kepler} observations are
incomplete for $R_P < 2R_{\oplus}$ and while the intrinsic radius
distribution of planets is presumed to turn over, the radius at which
this actually occurs is not known and awaits a better understanding of
the efficiency of {\it Kepler} detection of small signals.  If the
turnover below $2R_{\oplus}$ is real, then the intrinsic slope of the
radius distribution is {\it steeper} than observed ($\alpha = 2.6$).
But if a scale-free power-law distribution continues to much smaller
radii, then Eddington bias affects the magnitude, but not the slope of
the distribution.

\begin{figure}
\epsscale{0.7}
\plotone{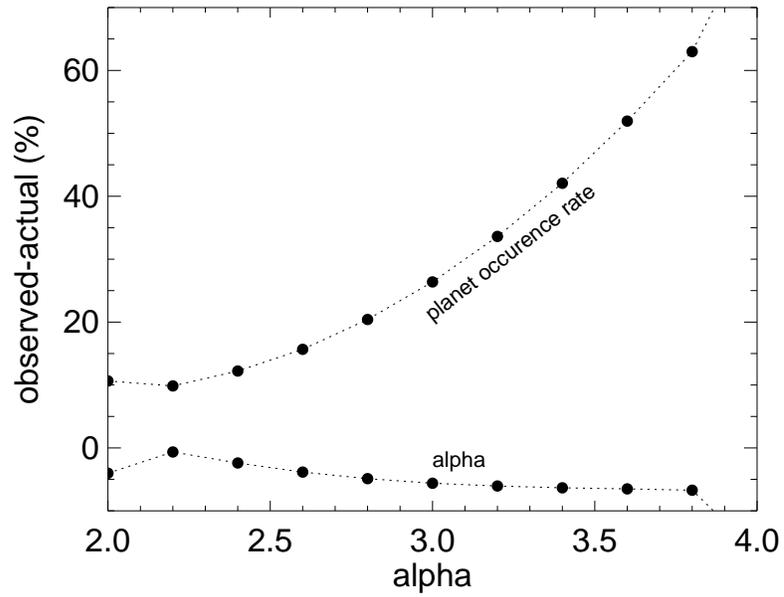}
\caption{Predicted biases in planet occurrence and power-law slope
  $\alpha$ due to Eddington bias for artifical planets with a
  power-law radius distribution placed around {\it Kepler} KOI-hosting
  stars.  \citet{Howard2012} report that $\alpha \approx 2.6$ for
  planets with periods $P< 50$~d.  The slope of the scale-free power
  law distribution is only slightly affected by Eddington bias, but
  the apparent occurrence is biased upwards because more numerous
  smaller planets appear as larger planets due to errors in stellar
  radius. \label{fig.eddington2}}
\end{figure}

We simulated Eddington bias on artificial samples of planets with
radii drawn from a power-law distribution with variable index
$\alpha$.  These radii replaced actual KOI estimates in a repeat of
the analysis described above.  The power-law index of the binned
apparent distribution $\rho(R_i)$ above some minimum radius $R_{min}$
is calculated by maximum likelihood: $\alpha = \sum_i \rho(R_i)/
\sum_i \rho(R_i) \log \left(R_i/R_{min}\right)$, where the summation
is over all $R_i > R_{min}$.  As expected, while Eddington bias
significantly increases the fraction of planets with $R > R_{min}$,
the power-law index is relatively unchanged (Figure
\ref{fig.eddington2}).

\subsection{Malmquist Bias \label{sec.malmquist}}

Malmquist bias is the preferential inclusion of intrinsically luminous
objects in a magnitude-limited survey due to the rapid increase in
sampling volume $d_{max}^3$ with distance $d_{max}$ to which an object
is included.  Among large, readily-detected objects (planets) in a
magnitude-limited transit survey, the bias is even greater ($\sim
d_{max}^4$) because the probability of a transiting geometry is
proprtional to $R_*$ which, at a given effective temperature, scales
with $d_{max}$ (see Section \ref{sec.targetbias}).  At a given
apparent magnitude and planet radius, there is a maximum stellar
radius $\tilde{R}_*$ to which a survey is essentially complete,
i.e. not limited by the SNR of a transit event.

We estimated $\tilde{R_*}$ as a function of $R_p$ by establishing the
detection limit at different {\it Kepler} magnitudes.  The {\it
  Kepler} target catalog was constructed with different criteria for
stars with $K_p < 14$ and $14 < K_p < 16$ \citep{Batalha2010}; it is
probably nearly complete for dwarf stars to $K_p = 14$ but only
includes selected dwarfs with $14 < K_p < 16$ \citep{Batalha2010}.  We
adopted a SNR limit of 7.1 and an observation period of 487~d (B12).
To estimate the noise of a typical dwarf star we performed a
polynomial fit to a running median ($N=1000$) of 3~hr combined
differential photometric precision (CDPP) values for {\it Kepler}
targets with $\log g > 4$, presumed mostly dwarfs.  This gave an
estimate of the intrinsic 3~hr RMS noise level as a function of $K_p$;
\begin{equation}
  \log \sigma_{3}({\rm dwarfs}) \approx -4.27 + 0.116(K_p-12) + 0.0247(K_p-12)^2.
\end{equation}
The median noise at $K_p = 12$ is 54 ppm.  We performed a similar
analysis on stars with KIC $\log g < 4$, presumably subgiants and
giants, that constitute a locus of comparatively ``noisy'' targets,
and found:
\begin{equation}
  \log \sigma_{3}({\rm giants}) \approx -3.69 + 0.045(K_p-12) + 0.115(K_p-12)^2.
\end{equation}
For $K_p = 14$ dwarfs, $\tilde{R_0} = 1.72R_{\odot}$ and at $K_p =
16$, $\tilde{R_0} = 0.77R_{\odot}$.  At $K_p = 14$, for a median
orbital period $P \approx 16$~d and $R_p = 2R_{\oplus}$, Malmquist
bias favors stars as large as $2.3R_{\odot}$.  At $K_P = 16$, only
stars with $R_* < 1.0R_{\odot}$ are favored because of higher noise at
fainter magnitudes.  The situation is more extreme for giant planets
($R_p \sim 10R_{\oplus}$), where Malmquist bias will favor evolved
stars as large as 10-20$R_{\odot}$, presuming giant planets exist
around such stars, as we discuss below.

Bias towards larger stars, coupled with uncertainties in stellar
radius, leads to underestimates of stellar - and hence planetary -
radii.  We quantified this effect using the machinery described in
Section \ref{sec.methods}, with the addition of a Malmquist bias
factor.  For each KOI-hosting star, we evaluated the mean stellar
radius by averaging over all stellar models weighted by $w$ (from
Equation \ref{eqn.weight}) and multipled by either
$(R_*/\tilde{R_*})^4$, where $R_* < \tilde{R}_*(P,K_p)$, or
$(R_*/\tilde{R}_*)^{-1/2}$, if $R_* < \tilde{R}_*(P,K_p)$.

\begin{figure}
\epsscale{0.7}
\plotone{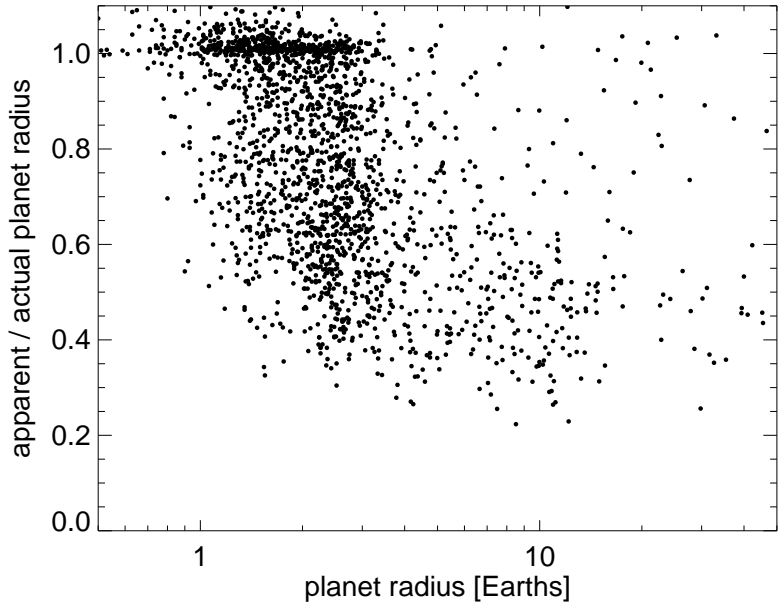}
\caption{Effect of Malmquist bias on the radii of stars and their
  planets.  The ratio of the apparent or ``naive'' radius to the
  actual or ``bias-informed'' radius of 2061 KOIs is plotted vs. the
  nominal planet radius from the catalog of \citet{Batalha2012}.  (239
  others are around stars with incomplete parameters from the Kepler
  Input Catalog).  The ``naive'' radius is the mean radius of possible
  stellar models weighted according to their consistency with KIC
  parameters and priors of mass, age, and metallicity.  The
  ``bias-informed'' radius is the mean calculated using the scaling
  laws for Malmquist bias derived in Section \ref{sec.targetbias}.
  1254 KOIs, and the vast majority of planet candidates smaller than
  Neptune, have predicted bias $<$10\%, but many giant ``planets'' may
  have radii twice the nominal value and some may be astrophysical
  false positives, i.e. eclipsing stars. \label{fig.malmquist}}
\end{figure}

The ratio of the ``naive'' mean model radius to the bias-weighted mean
radius is plotted in Figure \ref{fig.malmquist} vs. the nominal planet
radius published in B12.  Deviation of this factor from unity can be
considered the error in radius that results if Malmquist bias is not
taken into account.  About two-thirds of all KOI-hosting stars, and
the vast majority of those hosting planets smaller than Neptune have
predicted Malmquist bias values $<$10\%.  However, the majority of
larger planets may have significantly underestimated radii, some by a
factor of two.  This dichotomy occurs because {\it Kepler} detection
of large planets is limited by the magnitude limit of the target
catalog, not the SNR of transit.  We emphasize that these calculations
are {\it statistical}, i.e. we are calculating the expectation values
of probability distributions with stellar radius, and that actual
errors will vary.  Nevertheless, the host stars of many giant planets
may be more larger, more distant, and more luminous, and the radii of
their planets may be significantly underestimated.  Inclusion of
larger, evolved stars means that some KOIs may be astrophysical false
positives, e.g. M dwarf companions masquerading as planets
\citep{Charbonneau2004,Almenara2009}, a possibility that we discuss in
Section \ref{sec.discussion}.

\subsection{Metallicity Bias \label{sec.metallicity}}

The metallicity of host stars is an important parameter in studies of
planet statistics.  A correlation between stellar metallicity and the
presence of giant planets has been unambiguously established
\citep{Gonzalez1998,Santos2004,Fischer2005b,Buchhave2012} and is
consistent with a prediction by the core-triggered instability theory
of giant planet formation \citep{Mizuno1980}, i.e. that a solid core
that initiates runaway accretion before the gas dissipates is more
likely to form in a disk with a higher abundance of solids.  Doppler
surveys have failed to find any correlation between metallicity and
the occurrence of Neptune-size or smaller planets
\citep{Sousa2008,Bouchy2009, Mayor2011}.  \citet{Schlaufman2011} found
that the average $g$-$r$ color of most {\it Kepler} stars with small
candidate planets was no different from the average of all stars at a
given $J$-$H$ color, except for late K and early M-type stars; those
with planets have redder $g$-$r$ colors and
\citeauthor{Schlaufman2011} argued that these are more metal-rich.
However, this difference may be an artifact of contamination of the
sample by evolved stars, which have bluer $g$-$r$ colors than dwarfs
and make the overall sample, but not the KOI-hosting sample, bluer
\citep{Mann2012}.  Indeed, $g$-$r$ color might be insensitive to or
depend only weakly on metallicity for these spectral types
\citep{Lepine2012}.  \citet{Muirhead2012} report metallicities of 78
late K and M dwarfs with KOIs based on infrared spectra.  The mean
value, -0.09, is consistent with the metallicity of M dwarfs in the
solar neighborhood \citep{Laughlin2010,Woolf2012}.  The average
metallicity of {\it Kepler} M dwarfs is not known but these
intrinsically faint stars are within a few hundred pc of the Sun
\citep{Gaidos2012}.

The metallicities of stars of transiting planets need not be
representative of the underlying population of planet-hosting stars.
Metals are an important source of opacity in the atmospheres of cool
stars, and, all else being equal, metal-poor dwarf stars should have
smaller radii.  A transiting planet will be more detectable around a
metal-poor subdwarf than a metal-rich dwarf star, and thus the host
stars of KOIs will be biased towards metal-poor representatives of the
overall population.  If sufficiently large, this bias could obfuscate
any intrinsic relationship between stellar metallicity and the
presence of planets.

\begin{figure}
\epsscale{0.7}
\plotone{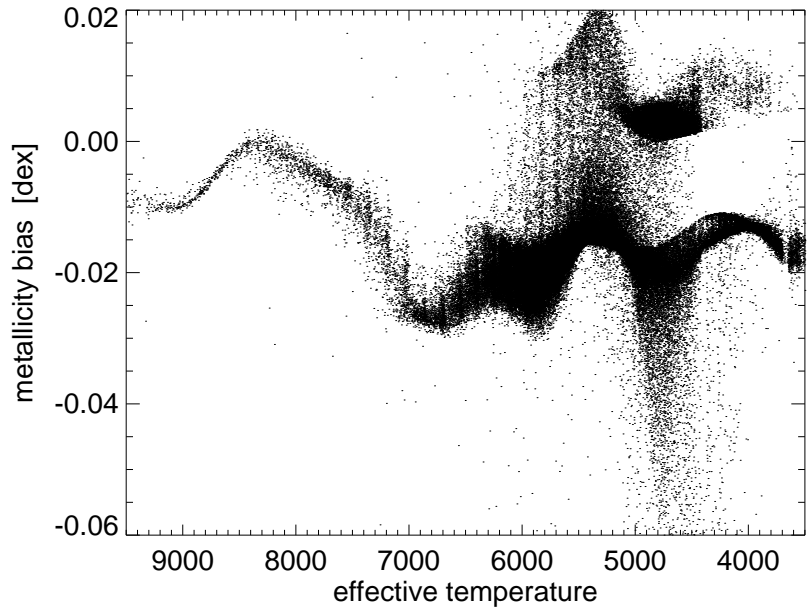}
\caption{Predicted bias in metalllicity in transiting planet-selected
  stars as a consequence of the relationships between transit depth,
  stellar radius, and stellar metallicity.  The bias was calculated
  for all Quarter 6 {\it Kepler} target stars (regardless of whether
  or not they host KOIs) using Eqns. \ref{eqn.bias} and
  \ref{eqn.occur} and the methods described in Section
  \ref{sec.methods}. The upper locus of positive values are evolved
  stars, for which radius {\it decreases} with increasing
  metallicity.  \label{fig.metalbias}}
\end{figure}

We calculated the metallicity bias, i.e the expected metallicity of
stars with detected transiting planets minus the expected metallicity,
for all {\it Kepler} Quarter 6 target stars using Eqns. \ref{eqn.bias}
and \ref{eqn.occur}, and the methods described in Section
\ref{sec.methods}.  The difference between the ``naive'' mean
metallicity of Dartmouth models for each star, and the biased mean
using the factor of Equation \ref{eqn.occur}, is plotted vs. adjusted
KIC effective temperature in Figure \ref{fig.metalbias}.  As expected,
the metallicity bias is negative except for a locus of positive values
corresponding to evolved stars, for which radius {\it decreases} with
increasing metallicity, e.g. \citet{Zielinski2012}.  The bias is small
(mean of -0.017 among dwarfs) for the following reasons: (i) the
geometric transit probability is proportional to stellar radius and
thus increases with metallicity, countering the effect of metallicity
on transit depth; and (ii) the effect of metallicity on stellar radius
is most pronounced among comparatively rare subdwarfs but has only a
modest effect around solar metallicity, especially for the coolest
stars \citep{Boyajian2012}.

\subsection{ Covariant errors  and ``inflated'' Jupiters \label{sec.inflated}}

At the time the first exoplanet around a main sequence star was
confirmed, \citet{Guillot1996} realized that highly-irradiated giant
planets on close-in orbits may have anomalously large radii.  After
sufficient numbers of transiting giant planets were discovered, it
became apparent that some were ``inflated'' compared to theoretical
predictions \citep{Burrows2000,Baraffe2003}.  Planets larger than $R_J
\approx 1.2$ cannot be explained by conventional interior models of
gas giants and require an additional source of internal energy to
inflate them \citep{Fortney2010}.  Several non-exclusive explanations
for the requisite energy source have been put forward
\citep{Bodenheimer2001,Showman2002,Batygin2010}.  One important clue
is that planets experiencing higher irradiance or having higher
emitting temperature are more likely to be inflated.  Correlations
between equilibrium temperature and radius have been reported among
transiting giant planets discovered in ground-based surveys
\citep{Laughlin2011,Enoch2012}.  Among {\it Kepler} giant planet
candidates, inflation appears to occur only above an irradiance of
about $2\times10^{8}$~ergs~s$^{-1}$~cm$^{-2}$ \citep[hereafter
D11]{Demory2011a}.

Where information about stellar parameters is limited, spurious
correlations can appear if two supposedly independent planetary
parameters are related to the same, uncertain stellar parameter.  In
the absence of parallax or precise information on surface gravity, the
radius of a star is constrained only by models of stellar atmospheres,
stellar evolution, and galactic population.  Uncertainty in stellar
radius translates into corresponding uncertainties in both stellar
luminosity and transiting planet radius.  Because the radiation that a
planet receives from a star is proportional to stellar luminosity,
errors in irradiance and planet radius due to errors in stellar radius
will be positively covariant.  At least in principle, an apparent,
positive trend between irradiation and planet radius could be created
merely by errors in stellar radius.

\begin{figure}
\epsscale{0.7}
\plotone{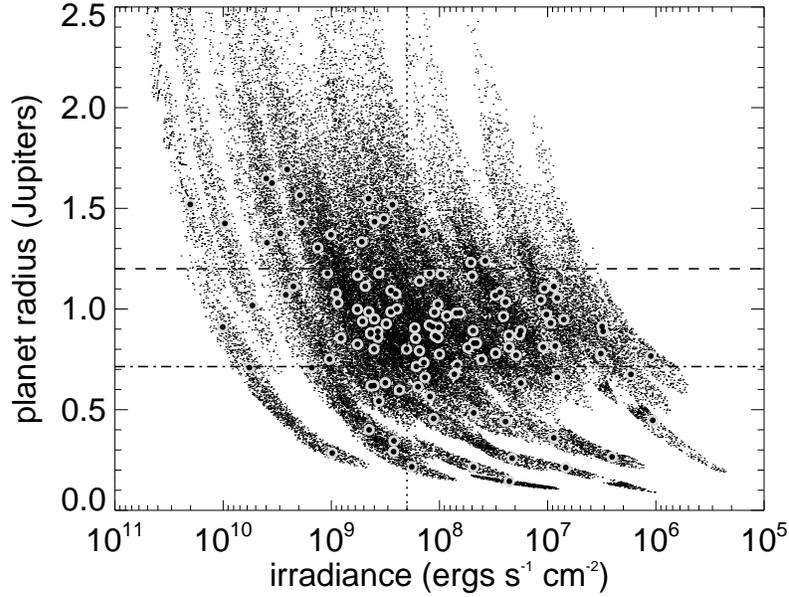}
\caption{Radius vs. stellar irradiance of candidate giant planets
  (8$R_{\oplus} < R_P < 22R_{\oplus}$) in the latest KOI release.
  These exclude the KOIs listed as false positives or ``ambiguous'' in
  the Table 1 of \citet{Demory2011a}.  Each large point represents
  values based on the stellar radius and luminosity of the Dartmouth
  stellar model that best reproduces the stellar parameters from the
  KIC.  The dots represent 200 models chosen randomly from among all
  Dartmouth stellar models that cannot be ruled out at 95.4\%
  ($2\sigma$) confidence.  The vertical dotted line demarks the
  suggested boundary between high and low stellar irradiation regimes.
  Objects above the horizontal dashed line (1.2$R_J$) are considered
  ``inflated''.  Objects below the dot-dashed line ($8R_{\oplus}$) are
  smaller than reported in the KOI catalog and may have problematic
  {\it Kepler} lightcurve analyses.  These were not included in the
  statistical analysis.  \label{fig.radirr}}
\end{figure}

We simulated the impact of this systematic with an analysis of KOIs
similar to, but not identical to that of D11.  We selected all KOIs
with estimated radii of $8R_{\oplus} < R_p < 22R_{\oplus}$ from B12,
excluding those listed as false positives or ``ambiguous'' in Table 1
of D11.  As in Section \ref{sec.methods}, we identified the best-fit
Dartmouth model for each host star based on a $\chi^2$ minimization of
the difference with adjusted KIC values of $T_e$, $\log g$, and
[Fe/H], after applying corrections of -100~K to $T_e$ and 0.17~dex to
[Fe/H] (Br11).  We assumed standard deviations of 200~K, 0.36~dex, and
0.3~dex, respectively based on Br11 and 190 stars where both KIC and
spectroscopy-based parameters are available (B12).  If no KIC value
for [Fe/H] was available we assumed solar metallicity.  To estimate
the maximum possible effect, no constraints other than the Dartmouth
evolutionary tracks were used, i.e. we equally weighted masses, ages
between 1-13~Gyr, and metallicities between -2.5 and +0.5 dex.
Orbit-averaged stellar irradition of the planet was calculated based
on the model luminosity and mass, the orbital period, and assuming a
circular orbit.  (Non-circular orbits change the mean irradiance only
slightly.)  Planet radius was calculated from the transit depth and
stellar model radius and we did not account for limb darkening.  The
encircled points in Figure \ref{fig.radirr} indicate the best-fit
planet radius vs. irradiance.  Three KOIs (217.01, 774.01, and
1547.01) have estimated irradiances $<2 \times 10^8$
ergs~s$^{-1}$~cm$^{-2}$ and $R_p > 1.2R_J$, but only marginally so.

Fifteen KOIs have re-estimated radii $<0.5R_J$ even though the values
listed in B12 exceed the criterion $>0.714R_J$.  Twelve of these have
KIC impact parameters $b > 1$, suggesting problematic (or extreme
grazing) transit solutions.  Another (KOI 1419.01) has an implausible
$b = 0.994$ which is inconsistent with its transit duration of $t =
1.36$~h and period $P = 1.36$~d.  KOI 377.02 ({\it Kepler} 9-b) has an
erroneous transit depth reported in the MAST.  The best-fit Dartmouth
model assigns a somewhat smaller radius ($0.48R_{\odot}$) to the host
star of KOI 1193.01 and thus makes the planet smaller as well.  We
excluded all planets with newly estimated radii $R_p < 8R_{\oplus}$
from our analysis.

We assessed the trends produced by correlated errors in planet radius
and irradiation by considering all Dartmouth models that satisfy
$\chi^2 < \chi_{min}^2 + 8.02$, where $\chi_{min}^2$ is the minimum
(best-fit) value, and 8.02 is the $\Delta \chi^2$ corresponding to a
95.4\% (2$\sigma$) confidence interval for $\nu = 3$ degrees of
freedom (stellar parameters).  Because there are too many models to
plot, we only show a random subsample of 200 such models for each KOI
as the small points in Figure \ref{fig.radirr}.  These clearly show that
correlated errors will tend to scatter points between the high
irradiation/inflated and the low irradiation/uninflated regions of the
diagram.

The paucity of KOIs with inflated radii ($R_p > 1.2R_J$) in the low
irradiance region (upper right hand domain of Figure \ref{fig.radirr})
supports the contention that the inflation of giant planets is related
to stellar irradiation or planet equilibrium temperature.
Furthermore, Kendall's and Spearman's rank correlation tests of all
KOIs with $R_p > 0.714R_J$ yield $\tau$ values of 0.246 and 0.364,
respectively, and corresponding $p$ (significance) values of $4.6
\times 10^{-5}$ and $3.1 \times 10^{-5}$.  These low false-positive
probabilities indicate a significant correlation between irradiation
and plane radius.  However, these statistics do not account for the
systematic effect of correlated errors in radius and irradiation.

\begin{figure}
\epsscale{0.7}
\plotone{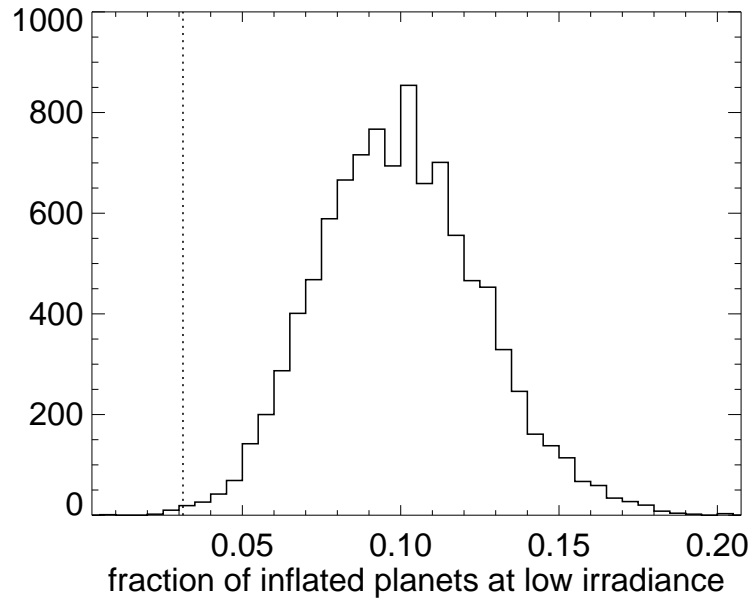}
\caption{Fraction of ``inflated'' ($R_p > 1.2R_J$) candidate planets
  in the low irradiance ($<$2$ \times 10^8$~ergs~sec$^{-1}$~cm$^{-2}$)
  regime in 1000 Monte Carlo simulations of the KOI data set where
  planets were randomly shuffled among stars and stellar parameters
  were resampled according to standard errors in the KIC values.  All
  KOIs with $R_p > 3R_{\oplus}$ were used to generate the artificial
  planet populations, but only planets with $R_p > 8R_{\oplus}$ were
  used in the analysis.  The vertical dashed line marks the actual
  number (3).  The $p$ value based on this distribution is $1.4 \times
  10^{-3}$.  \label{fig.ninflate}}
\end{figure}

\begin{figure}
\epsscale{0.7}
\plotone{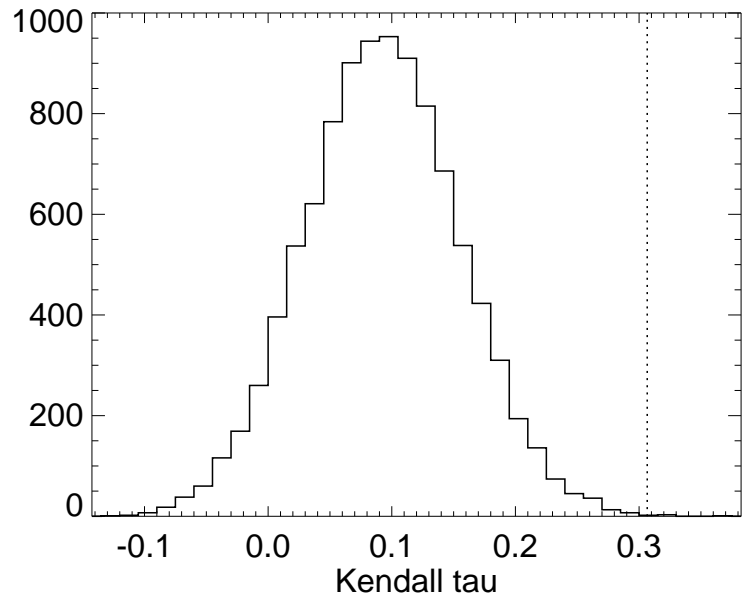}
\caption{Distribution of 1000 Kendall $\tau$ values for the
  correlation between planet radius and stellar irradiance using the
  same Monte Carlo realiziations of the giant planet KOIs as in
  Figure \ref{fig.ninflate}.  The vertical dashed line marks the actual
  value ($\tau = 0.31$), corresponding to a significane ($p$ value) of
  $6 \times 10^{-4}$.\label{fig.kendall}}
\end{figure}

We simulated the effect of correlated errors by analyzing 10000 null
realizations of the data where radii and orbital periods of KOIs were
randomly shuffled among host stars and the transit depths were
recomputed using Equation \ref{eqn.depth}, thus destroying any
intrinsic correlation between radius and irradiation.  In computing
each realization we include all KOIs with $R_p > 3R_{\oplus}$ to
account for small planets that may appear larger, but in each Monte
Carlo realization, as with the real sample, we limited the statistical
analysis to 8-22$R_{\oplus}$.  New (``observed'') estimates of KIC
stellar parameters were constructed from the ``true'' values by adding
random, gaussian-distributed offsets with standard deviations of 200~K
for $T_e$, 0.36~dex for $\log g$, and 0.3~dex for [Fe/H].  Best-fit
Dartmouth models were found for each parameter set, the planet radii
and irradiation values were determined, and the correlation statistics
were calculated.  New $p$ values for the fraction of KOIs in the
low-irradiance/inflated-radius zone, and Kendall's $\tau$, and
Spearman's $\tau$ were computed as the fraction of MC realizations
that are smaller (more significant) than the observed values.  The
distributions for the first two metrics are shown in
Figs. \ref{fig.ninflate} and \ref{fig.kendall} and the $p$ values are
$1.4 \times 10^{-3}$ and $6 \times 10^{-4}$, respectively.  The result
for the Spearman's rank coefficient is similar, with $p = 6 \times
10^{-4}$.

\subsection{Stellar metallicity and ``shrunken'' Jupiters \label{sec.shrunken}}

\citet[hereafter DR12]{Dodson-Robinson2012} reported a weakly
significant ($p=0.02$ or $2.3\sigma$) trend of decreasing radius of
{\it Kepler} (candidate) giant planet with increasing metallicity of
the host star.  She examined the ratio $R_p/R_*$ of 218 KOIs from
\citet{Borucki2011} with estimated radii of 5-20 $R_{\oplus}$ and the
correlation with estimated values of [Fe/H] from the KIC.  She
interpreted the decline as evidence that giant planets around
metal-rich stars have larger solid cores and, for the same total
planet mass, smaller radii \citep{Guillot2005}.

\begin{figure}
\epsscale{0.7}
\plotone{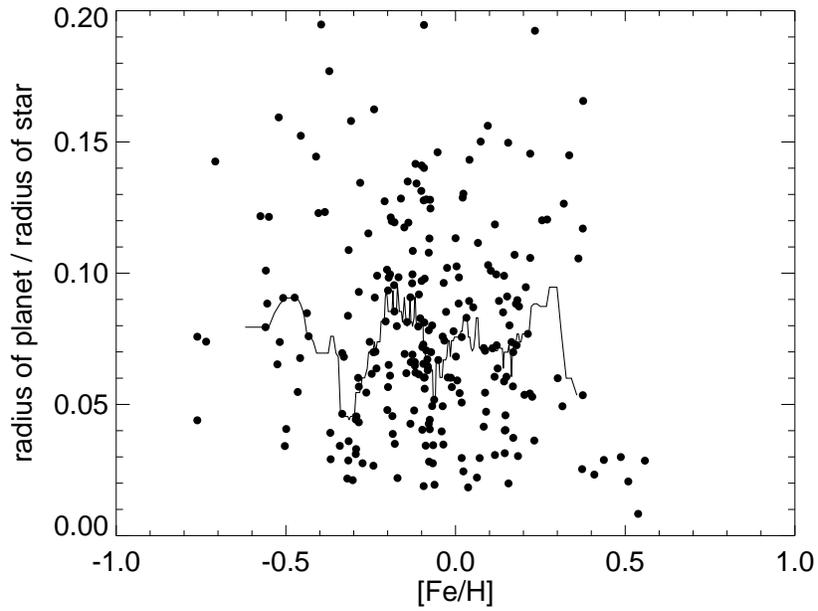}
\caption{Ratio of planet radius to star radius of 225 {\it Kepler}
  candidate planets with estimated radii between 5 and 20 $R_p$
  vs. metallicity estimates from the {\it Kepler} Input Catalog
  metallicities, uncertain by 0.3 dex (Br11). The curve is a running
  median (n=21).  No trend with metallicity is
  apparent.  \label{fig.depth}}
\end{figure}

Figure \ref{fig.depth} is an updated version of Figure 1 in DR12 based
on the more recent release of KOIs with revised radii
\citep{Batalha2012}.  It includes 225 KOIs with $5R_{\oplus} < R_p <
20R_{\oplus}$ and host stars with KIC-determined metallicities.  As in
Figure 1 from DR12, a running median ($N = 21$) is plotted.  The Kendall
$\tau$ correlation coefficient is -0.032, indicating no signficant
correlation ($p=0.48$). We were unable to reproduce the result of DR12
by simple cuts on this sample to approximate the earlier KOI sample,
perhaps because many stellar radii (and hence planet radii) have been
revised \citep{Batalha2012}.  We also emphasize that the values of
[Fe/H] in the KIC are no more accurate than $\pm$0.3 dex (Br11).

Irrespective of any physical phenomenon, one would expect to observe a
decrease in $R_p/R_*$ with increasing metallicity simply because
metal-rich dwarfs tend to be larger than metal-poor dwarfs, and hence
transit depths will be smaller (Equation \ref{eqn.depth}).  We modeled
this effect with 10000 Monte Carlo realizations of the KOI catalog.
There are two effects from increasing the radii of the host stars of a
given planet population: one is that transit depths will become
smaller and the planets will appear to be smaller.  The other is that
some planets may fall below the lower radius cutoff (5$R_{\oplus}$)
and be excluded from the analysis.  The reverse is true for lower
metallicity; planets appear larger and a few planets may exceed the
maximum cutoff (20$R_{\oplus}$).  We therefore considered KOIs over a
broader (3-25$R_{\oplus}$) range of radii, adopted this sample as
representing the intrinsic (``true'') distribution of radii, estimated
their apparent radii from the radius of the star and transit depth,
and then applied the same radius criteria as DR12.  We randomly
shuffled the planet population among the host stars, thus destroying
any intrinsic radius-metallicity correlation, computed the radii of
the stars using the Dartmouth stellar evolution models, and
re-calculated the transit depths.

Each Monte Carlo host star was assigned the corrected $T_e$ of the
actual star it replaced.  We assigned an {\it observed} metallicity
based on the KIC value, a systematic correction $\Delta$ of 0.17 dex
(Br11), a random normally-distributed error $\sigma$ of 0.3 dex, and a
prior distribution of {\it intrinsic} metallicities that is a guassian
with mean $\bar{F}$ and standard deviation $\epsilon$.  This is equivalent
to drawing metallicities from a single normal distribution with mean
$(\epsilon^2(\bar{F}+\Delta) + \sigma^2
\bar{F})/(\epsilon^2+\sigma^2)$ and standard deviation
$\epsilon^2\sigma^2/(\epsilon^2 + \sigma^2)$.  The radius of each
Monte Carlo star was taken to be the median of all model radii with
$\log g > 4$ (presuming they are dwarf stars), [Fe/H] within 0.15 dex
of the Monte Carlo model, and $T_e$ within 100~K.  We did not apply
any age criterion other than 1-13~Gyr.  We then calculated $R_p/R_*$
using the shuffled planet radius and the median model radius stellar
radius.  For each Monte Carlo sample, we calculated Kendall's $\tau$
and false positive probability for a correlation between the {\it
  observed} metallicities and the artificial transit depths.

\begin{figure}
\epsscale{0.7}
\plotone{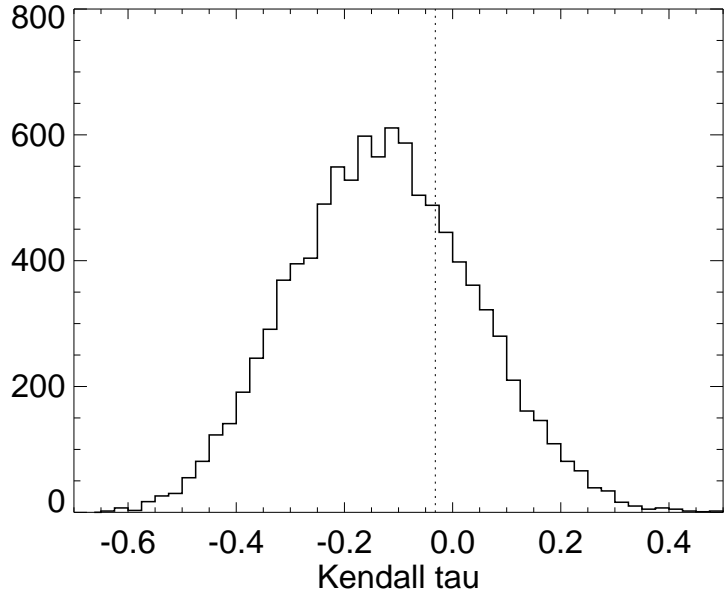}
\caption{Distribution of Kendall $\tau$ values among 10000 Monte Carlo
  simulations of the KOI data set shown in Figure \ref{fig.depth}.  Host
  star metallicities and KOI radii are scrambled, removing any
  intrinsic correlation between metallicity and planet radii.  The
  only correlation here is due to the increasing radius of stars with
  metallicity.  The dotted line is the value from the actual
  data. $3\sigma$ significance correspond to $\tau \approx -0.13$.
  Therefore, there is no significant correlation in the real data,
  and, because of the dependence of stellar radius on metallicity, a
  null sample would contain a significant ($p < 0.01$) correlation
  about 60\% of the time. \label{fig.tau}}
\end{figure}

Median-filtered ($n = 21$) curves from these Monte Carlo realizations
typically show a decline of $R_p/R_*$ with increasing metallicity.
Figure \ref{fig.tau} shows the distribution of $\tau$ from 10000 null
realizations.  The value of $\tau$ from the actual KOI sample is
plotted as the dashed line.  61.6\% of these null realizations produce
a significant ($p < 0.01$) correlation and 71.6\% of values are below
(and thus more significant than) the actual value of -0.032.  For
comparison the DR12 value is -0.17.  Thus, negative correlations
between metallicity and $R_p/R_*$ are to be expected soley as a
consequence of the metallicity-radius relation of stars, although
these Monte Carlo simulations indicate that there is a $\sim$40\%
chance that random errors in KIC [Fe/H] values would prevent such a
correlation from being detected.

\section{Discussion \label{sec.discussion}}

We have shown that selection effects for both transiting planets, and
the target stars of transit surveys, combined with uncertainties in
stellar radii, can bias the properties of host stars and their
planets.  These effects are in addition to those previously identified
by \citet{Gaudi2005}, \citet{Gaudi2005a}, and \citet{Pont2006a}, which
concern effects arising from the sensitivity of detection efficiency
to planet radius and period.  We have analyzed the effects of these
systematics on the {\it Kepler} survey and its catalogs of target
stars and candidate planets, using current models of stellar evolution
and galactic stellar populations to infer the properties of {\it
  Kepler} stars.  We did not apply constraints from the relation
between stellar density, transit duration, and orbital period because
the relation also depends on unknown orbital eccentricity and argument
of periastron, and is not applicable to non-KOI stars.

We found that Eddington bias from the steep distribution of KOIs with
radius results in an overestimation of the overall frequency of
planets with $R_p > 2R_{\oplus}$ by about 15-20\% of the actual value.
We also find that Eddington bias acts to soften the ``bump'' in the
distribution at Jupiter-size planets.  This leads us to predict that
the intrinsic peak at that radius is more pronounced.  The effect on
the distribution of smaller planets depends on whether the turnover in
the radius distribution below 2$R_{\oplus}$ is real, or the result of
incompleteness.  If the former, Eddington bias acts to flatten the
apparent slope of the radius distribution, and in this case we predict
that the actual slope is steeper than the $\alpha = 2.6$ power-law.
Otherwise, the effect of Eddington bias on the power-law index is
about 0.1.

We made statistical estimates of Malmquist bias as a consequence of
the magnitude limit of the target catalog.  The estimated bias for
two-thirds of KOI systems, including most KOIs smaller than Neptune,
is $<10$\%.  However, we found that bias is more prevalent and
pronounced (up to a factor of two in radius) among larger candidate
planets and their host stars, resulting from detection of these
systems being governed by the apparent magnitude limit of the target
catalog, rather than the SNR of transit detection.  A Malmquist bias
towards more luminous stars raises the possibility of inclusion of
unidentified evolved stars within the {\it Kepler} target catalog (in
addition to a number of deliberately selected and clearly identified
giant stars).  Nominally, stars with large radii were removed by a
vetting process that used a criterion of {\it Kepler} detection of a
$2R_{\oplus}$ planet \citep{Batalha2010}.  However, KIC-derived
stellar radii are based on estimates of log~$g$ and many of these are
problematic.  KIC photometry provides no information for the gravity
of stars with $T_e > 5400$K ($g-r < 0.65$), and subgiants would be
assigned erroneously high log~$g$ (Br11).

There are bona fida subgiants hosting KOIs, e.g. the F5 subgiant HD
179070 \citep{Howell2012}.  Spectroscopy of stars hosting candidate
giant planets has revealed other instances in which subgiants were
misclassified as cooler, main sequence dwarfs in the
KIC. \citet{Santerne2011} report a hot-Jupiter-hosting F-type subgiant
($M_* \approx 1.48M_{\odot}$, $R_* \approx 2.13R_{\odot}$).  Based on
spectra, they estimate $\log g = 4.1\pm0.2$, which is in contrast to
its KIC value of 4.55.  Likewise, the host of KOI-423b, assigned $\log
g = 4.45$ in the KIC, is an F7IV subgiant with $\log g = 4.1$
\citep{Bouchy2011}.  Three of five undiluted eclipsing binaries
identified by \citet{Santerne2012} as false positives among {\it
  Kepler} giant planet candidates have masses larger than 1 \msun, and
one of these is definitely an evolved star.  The mean difference
between 190 pairs of KIC and spectroscopic values of $\log g$ reported
in \citet{Batalha2012} is only 0.02~dex (standard deviation of
0.36~dex).  Nevertheless, astroseismically-derived $\log g$ values
average 0.05-0.17~dex lower than KIC values and
astroseismically-determined radii are up to 50\% larger
\citep{Verner2011,Bruntt2012}.

Among KOI-hosting stars whose radius has been underestimated, small
planets may actually be larger, even Jupiter-size planets.  In turn,
giant ``planets'' may turn out to be diluted or undiluted stellar
companions, a significant source of astrophysical false positives in
transit surveys \citep{Charbonneau2004,Almenara2009}.  Based on a
preliminary Doppler survey, \citet{Santerne2012} estimated that about
40\% of candidate giant planets are false positives and about one
quarter of those are undiluted eclipsing binaries.  This also means
that estimates of the occurrence of Jupiters on close-in orbits
\citep{Howard2012} must be revised downwards.  \citet{Wright2012}
report that the occurrence of ``hot Jupiters'' ($P < 10$~d) in the
{\it Kepler} catalog is only half that seen in Doppler surveys, and
adjustment for a high false-positive rate would worsen this
discrepancy.

One explanation for the discrepancy between the {\it Kepler} and
Doppler surveys might be the presence of misidentified subgiant stars
in the {\it Kepler} target catalog.  The intrinsic distribution of
planets may be different around evolved stars compared to main
sequence stars.  Planets have been discovered around subgiant stars
\citep{Butler2006}, but giant planets appear to be rare with 0.6~AU
($P < 120$~d) of clump GK giants \citep{Sato2008,Sato2010,Johnson2011}
- CoRoT-21b may be an exception \citep{Patzold2011}.  The timescale of
the decay of a planet's orbit due to dissipation of tides in a star's
convective envelope scales as $R_*^{-8}M_{env}$, where $M_{env}$ is
the mass of the envelope.  Hot Jupiters are likely to be destroyed by
infall and disruption inside the Roche lobe as a star evolves off the
main sequence, expands, and its convective envelopes thicken
\citep{Kunitomo2011}.  Thus, one explanation for the comparative
paucity of hot Jupiters in the KOI catalog is that, because of
Malmquist bias, many {\it Kepler} targets are older stars or subgiants
for which hot Jupiters cannot be detected, have been miscategorized as
Neptunes, or have been destroyed by orbital decay.  A comparison
between the distributions of $\log g$ predicted by TRILEGAL and that
of the KIC suggest no large ($>$10\%) population of unidentified
subgiants, however spectroscopy of candidate subgiants is needed to
actually test this conjecture.

We have shown that, because metal-poor stars tend to have smaller
radii than their metal-rich counterparts, stars with transiting
planets will be biased towards metal-poor members, independent of any
correlation between planets and metallicity.  However, we estimate
that this metallicity bias is only about -0.02 dex and can be
neglected.  Thus a comparison between the mean metallicity of stars
with transiting planets and that of the overall target population is
appropriate.  The mean metallicity of M dwarfs with KOIs, -0.09
\citep{Muirhead2012}, and solar-type stars with small planets, -0.01
\citep{Buchhave2012}, appears similar to that of the solar
neighborhood: \citet{Laughlin2010} report a mean metallicity of
$-0.14\pm0.06$ for a volume-limited local sample of M dwarfs using a
photometric calibration, and \citet{Casagrande2011} report a median
metallicity of -0.06 for all stars in the solar neighborhood.  Whether
the overall {\it Kepler} target population has a similar metallicity
distribution is not yet known and additional observations are
required.  From our calculations we conclude that such a comparison
would not suffer from significant metallicity bias, but must take into
account a dilution factor because stars without transiting planets are
not necessarily stars without planets.  This dilution factor is large
for a high planet occurrence \citep{Mann2012}.

We have shown how uncertainties in stellar radius or distance produce
correlated errors in a planet's radius and the radiation received from
the host star.  This effect can produce an artificial correlation in
populations of planets where none exists.  Recently, such a
correlation has been found in both ground-based transit surveys and
the {\it Kepler} catalog, and highlighted as a test of mechanisms to
explain the ``inflation'' of giant planets on close-in orbits.  We
quantified the systematic effect of correlated errors in stellar
radius in the case of the {\it Kepler} KOIs and show that, despite
this systematic, the result of D11, i.e. that inflated planets are
absent at low irradiance, is still significant.  To maximize any
systematic effect, we used a very broad range of metallicities (-2.5
to +0.5) and no constraint on stellar distance (e.g., from a model of
galactic structure), thus further strengthening our conclusion.

Finally, we have shown how searches for trends of transiting planet
radius with stellar properties may engender systematic errors unless
the effect of those properties on apparent stellar radius - and hence
planet radius - is taken into account.  We examined the tentative
(2.3$\sigma$) claim of DR12 that giant planets around metal-rich stars
tend to have smaller transit depths, because they are smaller and
perhaps have larger rocky cores.  Performing a similar analysis on the
most recent KOI catalog, we were unable to reproduce that trend.
Moreover, we performed simulations that show that the trend observed
by DR12 could be easily explained by the dependence of stellar radius
on metallicity.

Two limitations of our analysis are that (i) we have asssumed
gaussian-distributed errors in the corrected KIC parameters $T_e$,
$\log g$, and [Fe/H], and (ii) that the construction of Bayesian
priors on mass, age, and metallicity treat them as independent
variables.  Neither of these is absolutely correct; the first
assumption probably produces an underestimate of the uncertainty in
stellar radius while the second assumption produces an overestimate of
the uncertainty.  Of course, any inadequacies in the Dartmouth stellar
evolution models themselves are not accounted for.

There are other systematics effects which may be present in transit
surveys.  Two-thirds of solar-type (F6-K3) stars are found in multiple
systems \citep{Raghavan2010}.  At the typical distance of {\it Kepler}
KOIs with solar-type hosts (950~pc), one 4 arc-second pix subtends
about 3800~AU, sufficient to include nearly all companions to
primaries \citep{Lepine2007,Raghavan2010}.  The presence of an
unresolved companion, or any background star, will dilute the transit
signal.  Transits otherwise just above the detection threshold might
be rendered invisible.  As a consequence, members of multiple systems
may be underrepresented among stars with transiting planets.  For
equal-mass binaries (twins) where the transit signal is lower by a
factor of 2, the fractional noise will decrease by $\sqrt{2}$ (due to
the doubling of the signal compared to a comparable single star) and
thus the radius of the smallest detectable planet will increase by a
factor of $2^{1/4}$, or about 1.2.  For a power-law size distribution
(Equation \ref{eqn.distribution}), the number of detectable planets
{\it per star} will decrease by a factor of $2^{-\alpha/4}$, or 0.64
for $\alpha=2.6$.  However, nearly-equal mass binaries represent only
12\% of all binaries \citep{Raghavan2010} and systems with mass ratios
$< 1$ and luminosity ratios $\ll 1$, where the dilution will be much
smaller, are the norm.  Star counts reach
$\sim$1000~mag$^{-1}$~deg$^{-2}$ at $K_p = 16$, and so there is only a
few \% chance of significant dilution by an unrelated star.  To the
extent that stellar variability inhibits transit detection, younger,
and more active stars will be also underrepresented among KOIs.

The best defense against the systematic errors we have described is
better characterization of the target stars of transit surveys,
especially those hosting planets.  This will reduce, but not entirely
eliminate, these biases.  Spectroscopic characterization and
refinement of the properties of a fully representative sample of {\it
  Kepler} target stars, not just the KOI hosts, is vital to robust
statistical analyses of the properties of transiting planets and their
parent stars stars, and such programs are underway
\citep{Mann2012,Buchhave2012}.  Spectra of modest resolution ($R <
1000$) \citep{Malyuto2001} or SNR ($\sim 10$) \citep{Katz1998} (but
not both) can provide substantial improvements over photometry alone.
The {\it Gaia} (originally Global Astrometric Interferometer for
Astrophysics) mission, scheduled for launch in August 2013, will
obtain parallaxes of stars as faint as 16th magnitude with a standard
error of $\le$40 $\mu$as \citep{deBruijne2012}.  This will allow the
luminosity of a solar-type star to be determined with an error about
15\% and its radius with an error of about 8\%.  The distance to
brighter stars will be measured with even greater precision.  {\it
  Gaia} will also obtain moderate-resolution spectra in a narrow
region centered on the Ca II triplet region which can be used to
classify stars \citep{Kordopatis2011} and measure their radial
velocities to a precision of a few km~sec$^{-1}$.  Radial velocites,
combined with parallaxes, yield space motions and membership in
distinct stellar populations (e.g. thin disk, halo).  {\it Gaia} data
will also benefit future transit surveys that will cover all of or a
large part of the sky \citep{Deming2009}.

\acknowledgments

This research was supported by NSF grants AST-09-08406 and NASA grants
NNX10AI90G and NNX11AC33G to EG.  The {\it Kepler} mission is funded
by the NASA Science Mission Directorate, and data were obtained from
the Mukulski Archive at the Space Telescope Science Institute, funded
by NASA grant NNX09AF08G.

\clearpage

\begin{table}
\begin{center}
\caption{Parameter values for TRILEGAL 1.5 \label{tab.params}}
\begin{tabular}{ll}
\tableline
\tableline
Parameter & Value \\
\tableline
Dust: & \\
Extinction at $\infty$ & 0.0378 \\
Scale height & 110~pc \\
Scale length &  100~kpc \\
\tableline
Position of Sun: & \\
Galactocentric radius & 8700~pc \\
Height above disk & 24.2~pc \\
\tableline
Thin disk: & \\
Zero-age scale height & 95~pc\\
Radial length scale & 2.8~kpc\\
Local surface density & 59~M$_{\odot}$~pc$^{-2}$\\
Star formation rate & 2-step \\
\tableline
Thick disk: & \\
Scale height & 800~pc\\
Radial length scale & 2.8~kpc\\
Local density & $1.5\times 10^{-3}$~M$_{\odot}$~pc$^{-2}$\\
Star formation rate & 11-12~Gyr constant\\
\tableline
Halo: & \\
Shape & $r^{1/4}$ spheroid \\
Scale length & 2.8~kpc \\
Oblateness & 0.65 \\
Local density &  $1.5\times 10^{-4}$~M$_{\odot}$~pc$^{-2}$\\
Star formation rate & 12-13~Gyr\\
\tableline
\end{tabular}
\end{center}
\end{table}

\end{document}